\documentclass[twocolumn,aps,pra,floatfix,superscriptaddress]{revtex4-1}
\usepackage{amsmath,amssymb}
\usepackage{graphicx,float}
\usepackage{times,txfonts}
\usepackage{color}
\usepackage{multirow}
\usepackage{bbold}
\usepackage{color}
\usepackage{soul}
\usepackage{hyperref}
\usepackage{times,txfonts}
\usepackage{float}
\usepackage{natbib}
\usepackage{blindtext}
\usepackage[export]{adjustbox}
\usepackage{mathrsfs}
\usepackage{textcomp}
\usepackage{blkarray}
\usepackage{multirow}
\newcommand{\bra}[1]{\left\langle #1 \right|}
\newcommand{\ket}[1]{\left| #1 \right\rangle}

\renewcommand{\epsilon}{\varepsilon}

\def\VR{\kern-\arraycolsep\strut\vrule &\kern-\arraycolsep}
\def\vr{\kern-\arraycolsep & \kern-\arraycolsep}
\usepackage{sistyle}
\usepackage{soul}
\definecolor{lightblue}{RGB}{185,210,248}
\sethlcolor{lightblue}

\begin{document}
\title{Generation of doubly excited Rydberg states based on Rydberg antiblockade in a cold atomic ensemble}

\author{Jacob Taylor}
\affiliation{National Research Council of Canada, 100 Sussex Drive, Ottawa, Ontario K1A 0R6, Canada}
\affiliation{University of Waterloo, 200 University Avenue West, Waterloo, Ontario, N2L 3G1, Canada}
\author{Josiah Sinclair}
\affiliation{Department of Physics, University of Toronto, 60 St. George Street, Toronto, Ontario M5S 1A7, Canada}
\author{Kent Bonsma-Fisher}
\affiliation{National Research Council of Canada, 100 Sussex Drive, Ottawa, Ontario K1A 0R6, Canada}
\affiliation{Department of Physics, University of Toronto, 60 St. George Street, Toronto, Ontario M5S 1A7, Canada}
\author{Duncan England}
\affiliation{National Research Council of Canada, 100 Sussex Drive, Ottawa, Ontario K1A 0R6, Canada}
%\author{Aephraim Steinberg}
%\affiliation{Department of Physics, University of Toronto, 60 St. George Street, Toronto, Ontario M5S 1A7, Canada}
\author{Michael Spanner}
\author{Khabat Heshami}
\affiliation{National Research Council of Canada, 100 Sussex Drive, Ottawa, Ontario K1A 0R6, Canada}
\affiliation{Department of Physics, University of Ottawa, 25 Templeton Street, Ottawa, Ontario, K1N 6N5 Canada}

\begin{abstract}
Interaction between Rydberg atoms can significantly modify Rydberg excitation dynamics. Under a resonant driving field the Rydberg-Rydberg interaction in high-lying states can induce shifts in the atomic resonance such that a secondary Rydberg excitation becomes unlikely leading to the Rydberg blockade effect. In a related effect, off-resonant coupling of light to Rydberg states of atoms contributes to the Rydberg anti-blockade effect where the Rydberg interaction creates a resonant condition that promotes a secondary excitation in a Rydberg atomic gas. Here, we study the light-matter interaction and dynamics of off-resonant two-photon
excitations and include two- and three-atom Rydberg interactions and their
effect on excited state dynamics in an ensemble of cold atoms. In an
experimentally-motivated regime, we find the optimal physical parameters such
as Rabi frequencies, two-photon detuning, and pump duration to achieve
significant enhancement in the probability of generating doubly-excited
collective atomic states. This results in large auto-correlation values due to
the Rydberg anti-blockade effect and makes this system a potential candidate
for a high-purity two-photon Fock state source.
\end{abstract}

\maketitle
%%%%%%%%%%%
\section{Introduction}
Strong interactions between atoms excited to their high-lying Rydberg states
opens up many possibilities in quantum nonlinear optics \cite{ChangRev2014} and
atom-based quantum information processing \cite{SaffmanRev}. This stems from
the increased transition dipole moment between neighboring Rydberg states and
the stability (long lifetime) of Rydberg states due to the declining overlap between
Rydberg and ground electronic wavefunctions \cite{SibalicAdamsBook}. Mapping
photons onto Rydberg states of atoms enables implementation of photonic
entangling gates \cite{ParedesBarato2014,
Khazali2015,Saffman2015,Tiarks2019,Lukin2018,Ding2019}. The
most notable effect observed in optical excitation to Rydberg states is the
Rydberg blockade effect where one Rydberg atom shifts the energy levels of neighboring atoms such that they are no longer resonant with the driving field and no further atoms can be excited in this volume. Individually trapped atoms in high-lying 
Rydberg states together with the Rydberg blockade effect provides a platform for
simulation of many-body quantum dynamics \cite{Weimer2010, Bernien2017,
Keesling2019}. The Rydberg blockade effect in atomic ensembles allows one
to access a collectively-enhanced light-atom coupling with an effectively
isolated two-level system. This has applications in generating single photons
\cite{Montenegro2013, Craddock2019, Khazali2017}, quantum nonlinear optics effects
\cite{Peyronel2012}, strong interactions between photons \cite{Kuzmich2012,
Gorshkov2011, Hoferberth2014, He2014, Cantu2019, Sinclair2019}, and the development of quantum repeaters
\cite{Han2010}. 

The distance-dependent Rydberg-Rydberg interaction offers a playground to study
many-body dynamics and correlation of Rydberg excitations in atomic
ensembles~\cite{Stanojevic2010,Wuster2010}. Off-resonant driving of electronic
transitions to Rydberg states of atoms can result in the Rydberg anti-blockade
effect where the Rydberg interaction leads to preferential 2-atom excitation at a certain pairwise distance between atoms~\cite{Ates2007}. Under such
conditions, the Rydberg excitation dynamics can be significantly modified and
exhibit non-classical statistics. Measuring the ionization signal from a
Rydberg atomic gas has enabled experimental observation of the Rydberg antiblockade
effect~\cite{Amthor2010,Viteau2012}, however, important questions like the optimal parameter regime to observe
antiblockade, or whether it is possible to observe a signature of Rydberg antiblockade in transmitted
photon statistics, remains open.

Exact treatment of the many-body Rydberg excitation dynamics is intractable. A
perturbative approach to Rydberg excitation dynamics uncovers non-classical
correlations and the distance-dependence of these
correlations~\cite{Stanojevic2010}. Considering the collective states of
Rydberg atoms has explained experimental observation of many-body
correlations~\cite{Viteau2012}. However, these approaches are limited in
their capability of finding optimal experimental conditions or in including the effect of average
atomic distance (atomic density) to reach the Rydberg anti-blockade condition
needed to prepare high-purity excited states of the Rydberg gas. 

In this work,
we use exact numerical solutions to the Bloch equations of interacting two-atom
and three-atom systems sampled from many possible configurations in a gas of
about 1000 atoms in a dipole trap. This enables us to study Rydberg excitation
dynamics and find optimal values for experimental parameters that maximize
two-atom excitation probabilities accompanied by large auto-correlation values
resulting in high-purity states.  In the remainder of the manuscript, we first
describe our scheme and model; see sections~\ref{scheme} and \ref{Model}. In section~\ref{results}, we demonstrate our results for two-photon excitation of Rydberg atoms under conditions similar to the Raman transition or Electromagnetically Induced Transparency (EIT). Finally, we show optimal conditions such as pulse durations, detuning, and single-atom Rabi frequencies and their effect on two- and three-atom excitation probabilities and the Rydberg-Rydberg correlations. Our approach paves the way to engineer Rydberg atom ensembles for generating non-classical states of light, such as Fock states of $n=2$. This approach could be optimized to reach low order Fock states of $n=3$, or $n=4$.

\section{Scheme}\label{scheme}
\begin{figure}[ht]
\includegraphics*[viewport=50 287 450 700,width=\linewidth]{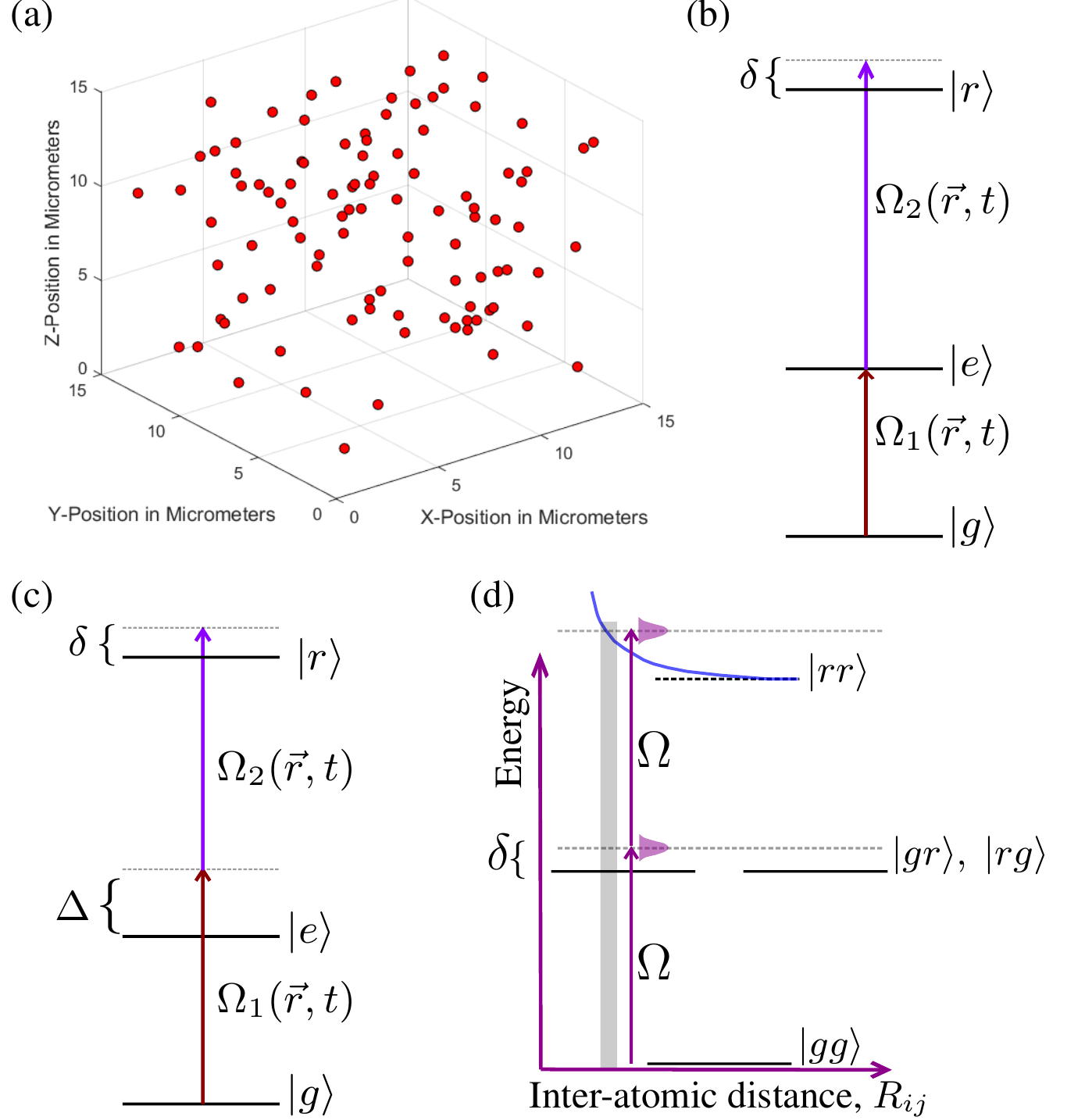}
\caption{(Color online) Scheme. (a) An example of a cloud of atoms at random
positions. In our simulations we take averages of the probabilities for pairwise
interactions for 1000 atoms randomly placed within a $14\mu m$ x $14\mu m$ x
$14\mu m$ box. (b) and (c) show schematic diagrams of resonant coupling
resembling the conditions for electromagnetically induced transparency, and
off-resonant Raman transition. (d) Schematic diagram of the effective 3 level system shows how the Rydberg-Rydberg interaction potential brings the
$\ket{rr}$ level into resonance at certain distances within the shaded area. For each single Rydberg atom the first excitation energy is out of resonance by $\delta$, yet the $\ket{rr}$ state can
be brought closer to resonance at $|2\delta - V_{ij}|\rightarrow 0$, where $V_{ij}$ is the Rydberg interaction between atom $i$ and $j$.}
\label{fig:fig1}
\end{figure}

We consider a gas of ultracold Rb atoms in a dipole trap and aim to address the
{\it nS} Rydberg states via a two-photon transition; see Fig.~\ref{fig:fig1}. This can be
achieved by resonant (Fig.~\ref{fig:fig1}(b)), or off-resonant (Fig.~\ref{fig:fig1}(c)) coupling to an
intermediate electronic state of Rb atoms. We refer to these cases as electromagnetically induced transparency (EIT) and Raman-type coupling, respectively.
Atoms excited to their Rydberg
states interact with neighboring atoms at moderate distances through the Van
der Waals interaction~\cite{SaffmanRev,SibalicAdamsBook}. Simplifying a pair of Rydberg atoms to the two ground and two Rydberg states driven by an effective
coupling, $\Omega(t)$, allows us to consider a two-atom system with four states
that include a global ground state, two degenerate single-atom excitations, and
a doubly excited Rydberg state; see Fig.~\ref{fig:fig1}(d). As it is shown in Fig.~\ref{fig:fig1}(d), single
Rydberg excitations are off-resonant with respect to the effective (ground to
Rydberg) coupling. Nevertheless, there is a non-zero probability to generate a
single-atom excitation through off-resonant scattering. In a non-interacting
system, the probability of generating a secondary excited atom scales as
the square of the probability of generating one excited atom. However, at a
certain inter-atomic distance, the Van der Waals interaction can generate a
resonant condition such that doubly-excited states of the two-atom system are
more likely to be reached. This satisfies the Rydberg anti-blockade conditions
which will depend on laser detunings, intensity, and Rydberg interaction
strength at the average inter-atomic distances in the atomic gas.

\section{Model}\label{Model}
\begin{figure}[ht]
\includegraphics*[viewport =50 490 550 700,width=1\linewidth]{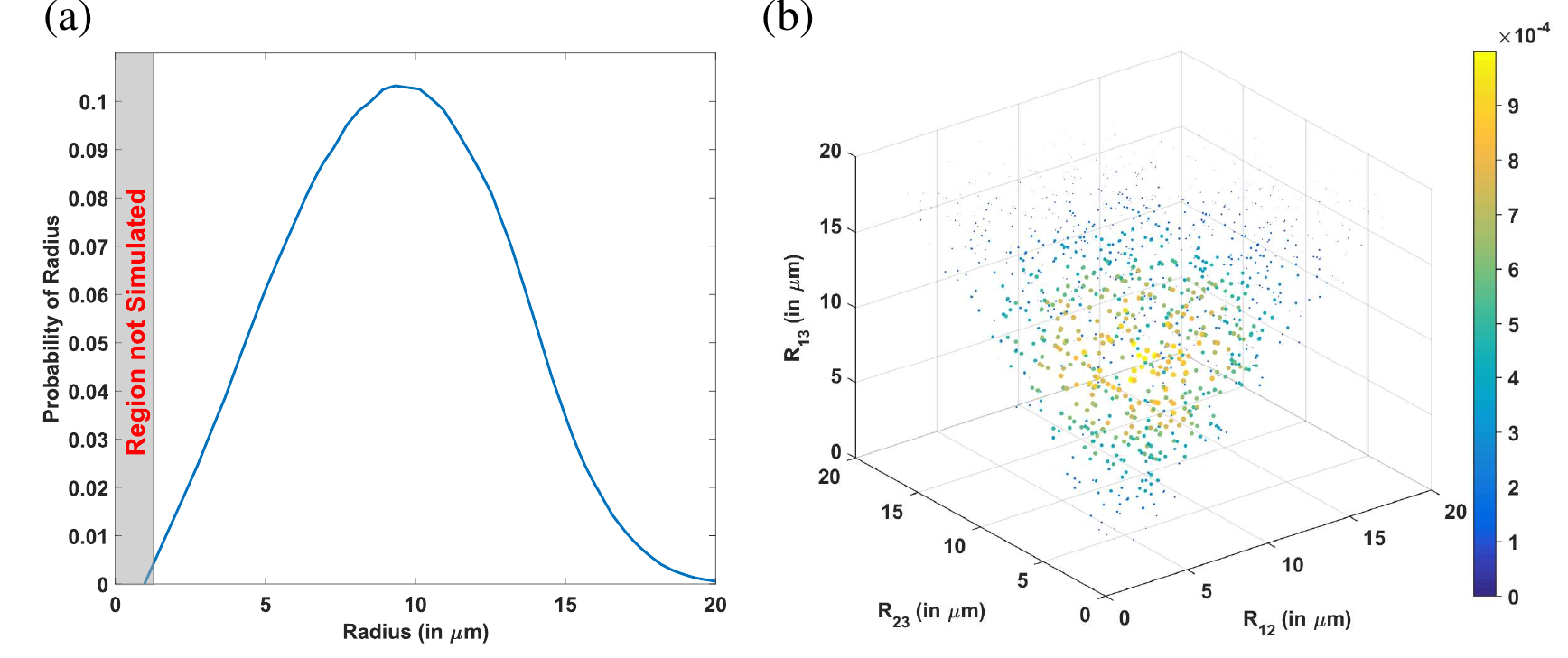}
\caption{(Color online) Probability density distributions w.r.t. pairwise atomic distance. (a) Shows the probability
density function for the distance between two randomly selected atoms within a
box of $14\mu m \times 14\mu m\times 14\mu m$ occupied by 1000 atoms. The
shaded region does not have a significant probability and is ignored for our
simulations to avoid numerical difficulties associated with large Rydberg interaction values. (b) Demonstrates the probability density function for three atoms, where the color and size of each point is used to represent the $P_{DF}$ of 3 randomly placed atoms having a triangle connecting them of
$R_{12},R_{23},R_{13}$ side lengths. In both cases the distribution peaks
around 10 microns. This is used in Sec.~\ref{results}(C) to study 3-atom Rydberg excitation dynamics.}
\label{fig:fig2}
\end{figure}

Since the Rydberg-Rydberg interaction strength varies significantly with
distance between every two atoms, it is essential to include these variations
in finding the optimal condition for the Rydberg anti-blockade effect. To this end,
we place 1000 atoms at random positions in a cloud of $14\mu m\times 14\mu m \times 14\mu m$, as shown in Fig.~\ref{fig:fig1}(a). Note that the density of the
gas follows an experimentally-relevant distribution in space. We take
two distinct approaches to considering the varying effect of atom-atom
interaction. The first approach is to randomly sample from the atomic cloud, calculate the
Rydberg excitation dynamics for each instance, and average over the resulting
probabilities. The accuracy of this approach increases by increasing the number
of instances. The second approach, on the other hand, extracts a probability
density distribution for pairwise distances of two-atom (Fig.~\ref{fig:fig2}(a)) and
three-atom (Fig.~\ref{fig:fig2}(b)) configurations. We can then use these probability
density distributions to average over Rydberg excitation dynamics outcomes of
all possible two-atom and three-atom configurations in the cloud. To
calculate the probability of a two-atom excitation, $P_2$ for example, with
$P_{DF}(R)$ being the probability distribution function of pairwise distance between two atoms (shown in Fig.~\ref{fig:fig2}), we use $P_2=\int P_2(R)*P_{DF}(R) dR$. Here, $P_2(R)$ is the probability of two atoms being excited to their Rydberg states at a distance $R$.
This can be extended to 3 atoms,
with a probability density function of three pairwise distances. In this case
$P_2$ is given by
\begin{equation}\label{PDF}
P_2=\iiint P_2(R_{12},R_{23},R_{13}) P_{DF}(R_{12},R_{23},R_{13})dR_{12}dR_{23}dR_{13},
\end{equation}
which will be used in Sec.~\ref{results}(C) for three-atom Rydberg excitation dynamics.

We take a semi-classical approach to study the Rydberg excitation dynamics in
the proposed system and treat the optical fields used to excite to $nS$ states 
classically. This is justified as both of these field are assumed to be
bright laser pulses. Within the rotating reference frame the multi-atom
Hamiltonian can be formulated as 
\begin{multline}
H=\sum_{i}\frac{\Omega_1}{2}\ket{g}\bra{e}_i+h.c. + \sum_{i}\frac{\Omega_2}{2}\ket{e}\bra{r}_i+h.c. -\delta\sum_{i}\ket{r}\bra{r}_i \\ -\Delta\sum_{i}\ket{e}\bra{e}_i+\sum_{j>i} V_{ij}\ket{r}_j\ket{r}_i\bra{r}_i\bra{r}_j .
\end{multline}
The Hamiltonian includes the Rabi oscillations that occur between $\ket{g}$,
$\ket{e}$, the ground and intermediate states, and $\ket{e}$ and the Rydberg
state $\ket{r}$ on each atom. The detuning from $\ket{r}$ and $\ket{e}$ levels
are determined by the pump lasers. The last term in the Hamiltonian describes
the Van der Waals interaction between two Rydberg atoms and is given by
$V_{ij}=\frac{C_6}{R_{ij}^6}$, where $C_6$ is the Rydberg interaction
coefficient associated with the Rydberg states we are addressing and $R_{ij}$ is
the distance between atoms $i$ and $j$. For two atoms, this can be simplified
to $H=H_\Omega+H_\delta+H_{int}$ where,
$H_\Omega=\frac{\Omega_1}{2}\ket{g}\bra{e}_1+\frac{\Omega_1}{2}\ket{g}\bra{e}_2+\frac{\Omega_2}{2}\ket{e}\bra{r}_1+\frac{\Omega_2}{2}\ket{e}\bra{r}_2+h.c.$,
$H_\delta=-\delta\ket{r}\bra{r}_1-\delta\ket{r}\bra{r}_2-\Delta\ket{e}\bra{e}_1-\Delta\ket{e}\bra{e}_2$,
and $H_{int}=V_{12}\ket{r}_2\ket{r}_1\bra{r}_1\bra{r}_2$. 
To simulate the Rydberg excitation dynamics we numerically evaluate the
time evolution of the density matrix using the master equation;
$\dot{\rho}=-\frac{i}{\hbar}[H,\rho]-\frac{\gamma}{2}\sum_{i}
\sigma_i^\dag\sigma_i \rho+\rho\sigma_i\sigma_i^\dag
-2\sigma_i\rho\sigma_i^\dag$, where $\sigma_i=\ket{g}\bra{r}_i$. We use the same formalism to account for decay from the intermediate state, $\ket{e}$.
The Rabi frequencies, $\Omega_1$ and $\Omega_2$,
can vary both in position and time to account for the optical pulse envelope
and a spatially-varying intensity. The Rabi frequency for each
transition is given by $\Omega(r_0,t)=\frac{\mu*E(r_0,t)}{\hbar}$ where
$E(r_0,t)$ is the amplitude of the classical field at $r_0$ with respect to the
center of the beam, and $\mu$ is the transition dipole of each transition.

In order to evaluate the Rydberg excitation statistics we define the following
projection operator 
\begin{multline*}
Proj_n=\sum_{\forall \ket{g...e}\in {N-n}}\sum_{i_1>i_2>...>i_n\geq 1}^{N}\ket{g...e...g...e}_{N-n}\ket{r_{i_1}...r_{i_2}...r_{i_n}}_{n}\\ _{n}\bra{r_{i_1}...r_{i_2}...r_{i_n}}~_{N-n}\bra{g...e...g...e}
\end{multline*}
that selects a subset of states with $n$ atoms excited to their Rydberg states
out of $N$ atoms. This projection operator must include all combinations
of $N-n$ atoms in $\ket{g}$ and $\ket{e}$ states. For example in the two-atom case, for $P_1$ both $\ket{r}\ket{g}$
and $\ket{r}\ket{e}$ need to be considered. More explicitly
$Proj_1=\ket{r}\ket{g}\bra{g}\bra{r}+\ket{g}\ket{r}\bra{r}\bra{g}+\ket{e}\ket{r}\bra{r}\bra{e}+\ket{r}\ket{e}\bra{e}\bra{r}$. The expectation values of this projection operator gives the probability of having one and only one Rydberg excitation; $P_1=\langle Proj_1 \rangle$.

\section{Results}\label{results}
In the case of addressing non-interacting atoms, the probability of generating doubly-excited states scales quadratically with the probability of generating a single excitation in the ensemble. This arises from the point that generating one excitation has no impact on atomic excitation dynamics when the system is far from saturation. Therefore, observing variations in $P_2/P_1^2$ is an indicator for generation of non-classical statistics. 

Below, we show our numerical results for generating non-classical statistics in an atomic ensemble. We explicitly consider two- and three-atom Rydberg excitation dynamics and find the optimal experimental condition to observe non-classical statistics arising from the Rydberg anti-blockade effect.

We numerically study two- and three-atom Rydberg excitation dynamics where each atom undergoes a two-photon transition to the $75S$ state; see Figs.~\ref{fig:fig1}(b) and \ref{fig:fig1}(c). We separately find optimal conditions for two-photon transition under resonant and off-resonant conditions with respect to the intermediate state ($\ket{e}$) which corresponds to single-photon detuning of $\Delta=0$ and $\Delta\neq 0$, respectively. We refer to these conditions as electromagnetically induced transparency (EIT) and Raman-type coupling.

\subsection{Raman-type coupling}
For our simulations, we use Rabi frequencies ($\Omega_1$,$\Omega_2$) with
a Gaussian temporal envelope and a flat spatial profile,
\begin{equation}
\Omega_i(t)=A_i\exp(-\frac{4\ln(2)t^2}{T_{FWHM}^2}), \end{equation}
where $T_{FWHM}$ is the full width half max of the pulse in time domain, and $A_i$
determines the peak Rabi frequencies. For most of our results $R_{FWHM}$ is
taken to relatively large values to simulate a flat intensity distribution in the cross
section of the pump interacting with the ensemble. The Rydberg-Rydberg interaction is given by $V_{ij}=\frac{C_6}{R_{ij}^6}$, where $C_6=11.66\times10^6$ MHz$\times$ $(\mu m)^6$ for the $75S$ state.

\begin{figure}[ht]
\includegraphics*[viewport =50 492 570 710,width=1\linewidth]{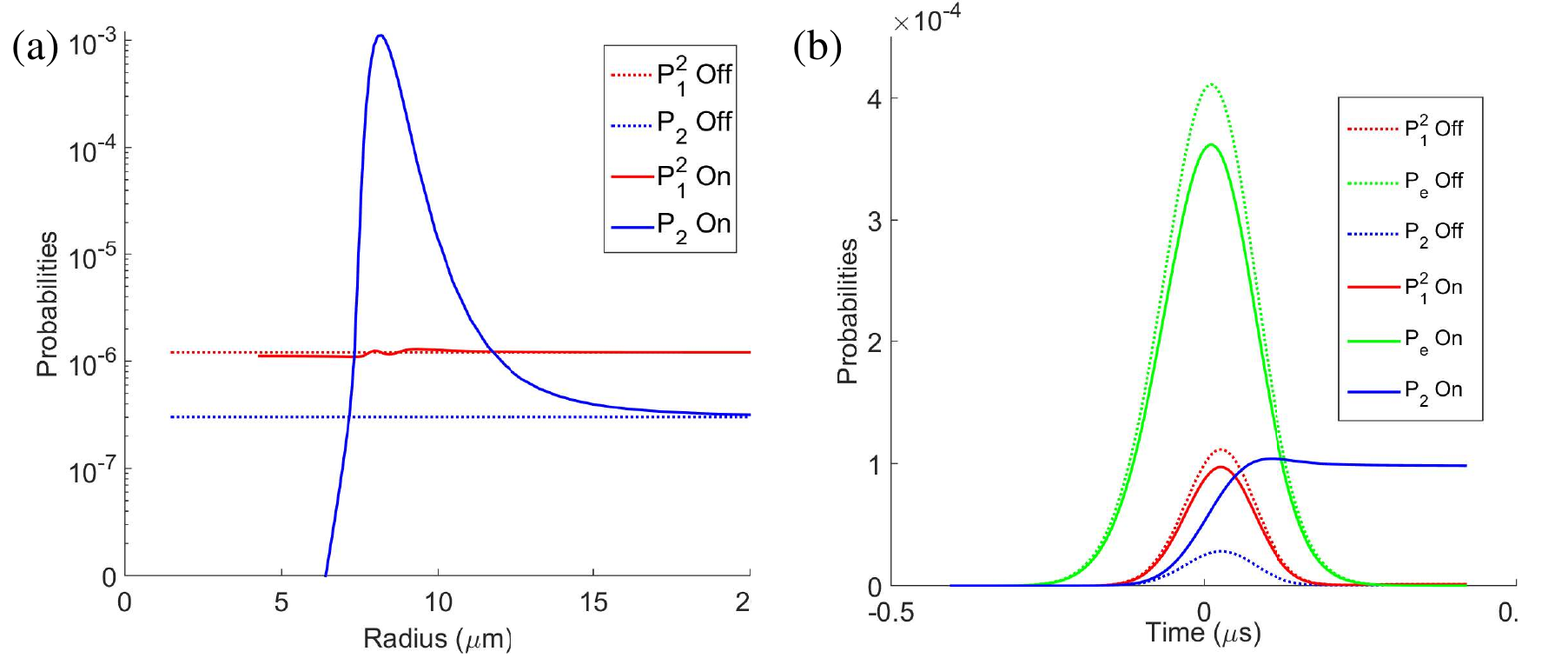}
	\caption{(Color online) Rydberg excitation dynamics . (a) Shows
probability of single and double Rydberg excitation in a two-atom system at
various distances where `On'/`Off' refers to the cases where the Rydberg-Rydberg interaction
is `included'/`not included'. In this example, $\delta=25\text{MHz}$,
$\Delta=100\text{MHz}$ $\Omega_1=18.6\text{MHz}$, $\Omega_2=30\text{MHz}$, and
the Rydberg level is 75S with a lifetime of $\frac{1}{\gamma}=179\mu s$. We
set the pulse duration to $T_{FWHM}=0.38\mu s$ and $R_{FWHM}=\infty$ as we
assume a flat pump in the transverse direction. The graph shows a significant deviation from the
non-interacting atomic systems, signifying the existence of the anti-blockade effect. This also amplifies the importance of setting the detuning, $\delta$,
such that to have as many atoms as possible within the anti-blockade region. (b) Shows an
example of Rydberg excitation dynamics when $T_{FWHM}$ is set to its optimal
value. Averaging over all possible atomic configurations results in
$P_2=0.0001$ and $P_2/P_1^2=94$.}
	\label{fig:fig3}
\end{figure}

Fig.~\ref{fig:fig3}(a) shows steady state values for $P_2$ and $P_1$ in a two-atom system at
various separations (5-20 microns) driven by two pulses. For small radius ($R
\lesssim 7\mu m$), $|2\delta-V_{ij}|$ becomes very large, thus $P_2$ approaches
zero.  This is the blockade regime, acting as expected. For slightly larger
$R$, the excitation probabilities reach a maximum around a resonant radius;
this is the anti-blockade regime.  As $R$ is increased far beyond the resonant
radius, the excitation probabilities slowly drop to their uncorrelated limits
as $V_{ij}$ goes to zero for large $R$.  As stated before whether blockade or anti-blockade occurs is determined by whether $|2\delta-V_{ij}|$ grows or
shrinks, with the optimal case occurring for $2\delta=V_{ij}$. For this optimal
case, the double excitation is in resonance, and this is the point for which the greatest value of $P_2/P_1^2$ is achieved; see also Fig.~\ref{fig:fig1}(d). Note that in case of non-interacting atoms $P_2\neq P_1^2$. This is due to our definition $P_1$ and $P_2$. As it can be seen in the definition of the projection operators $P_1$ is the probability of one and only one excited Rydberg atom. For small probabilities, one can show that $P_2/P_1^2$ is approximately $1/4$.

\begin{figure}[ht]
\includegraphics*[viewport =45 310 540 750,width=1\columnwidth]{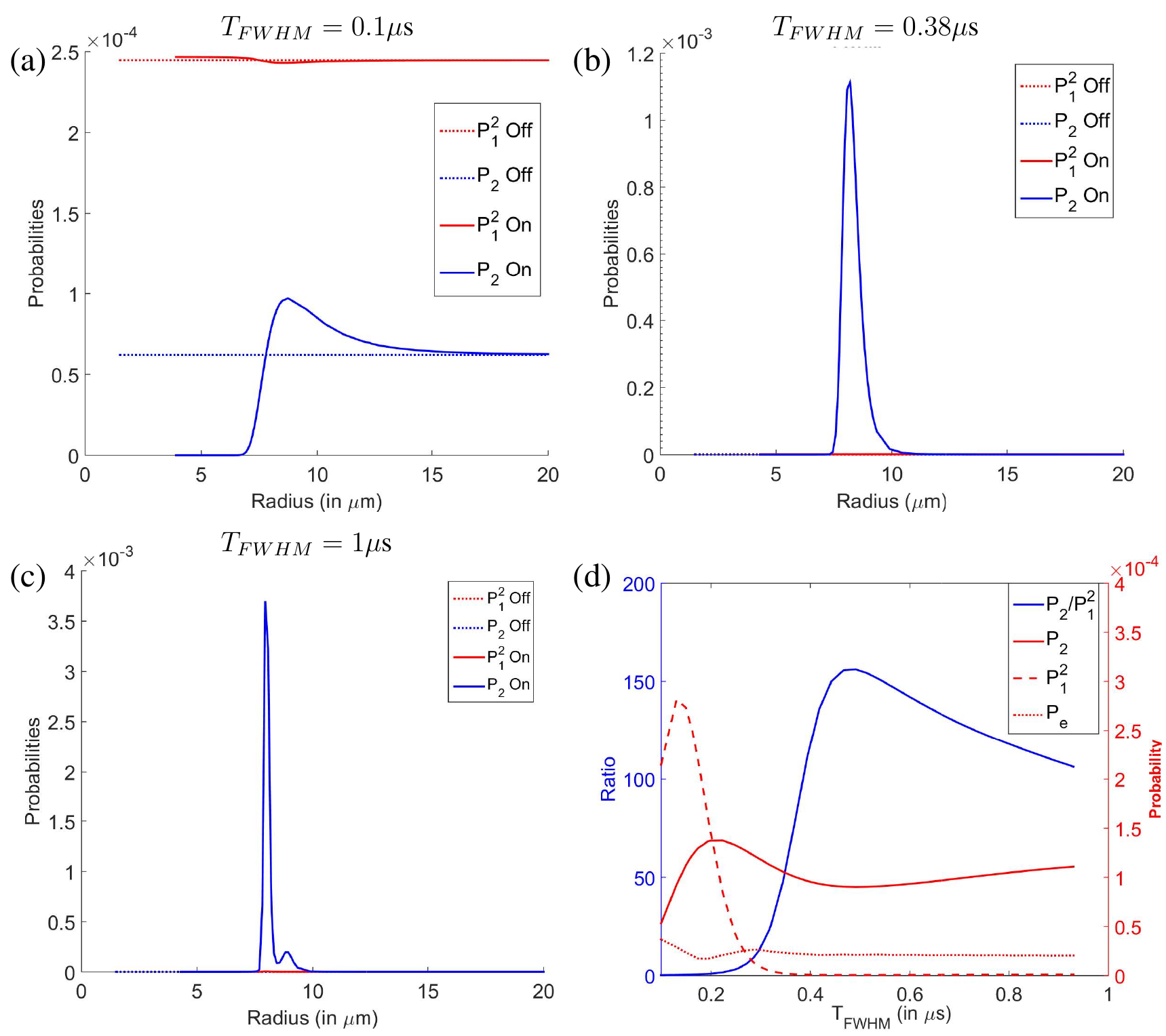}
\caption{(Color online) Rydberg excitation probabilities under different pump duration values. (a)--(c) Show two-atom Rydberg excitation prababilities for various atomic distances. All parameters are kept constant from Fig.~\ref{fig:fig3}(b) except for $T_{FWHM}$. Note that part (b) is identical to Fig.~\ref{fig:fig3}(a) on a linear scale axis. (d) Demonstrates the effect of $T_{FWHM}$ on the Ratio and Rydberg excitation probabilities. This graph uses the same physical parameters as in the
previous and each point is calculated by integrating relevant radius plots with the probability density distribution from Fig.~\ref{fig:fig2}.}
\label{fig:fig4}
\end{figure}

Parameters such as pulse duration and amplitude affect the Rydberg excitation
probabilities, its dynamics, and the purity of generating two-atom excited
states. Results in Fig.~\ref{fig:fig4} indicate that the $P_2/P_1^2$ grows with increase in pulse duration,
$T_{FWHM}$. The interaction strength dictates to what extent $P_2$ can be
increased without affecting the $P_2/P_1^2$ ratio. Modification of the
pump pulse duration influences two competing factors of the
anti-blockade when we consider the radius graphs (Fig.~\ref{fig:fig4}(a--c)). First, a larger
$T_{FWHM}$ yields a narrower pump beam in the frequency domain, and thus less
overlap between energy levels that are not exactly at resonance. This is
exemplified by a decrease in the width of the optimal inter-atomic distances
that show the anti-blockade effect in Figs.~\ref{fig:fig4}(a--c). Second, peak value of the excitation
probability at resonance increases with increasing pulse area of the $\Omega$,
and hence increases for larger values of $T_{FWHM}$. We find the optimal pulse duration that maximizes the $P_2/P_1^2$ ratio to occur around $0.5~\mu s$. For a fixed ratio of about 100, a pulse duration of $0.38~\mu s$ maximizes the $P_2$; see Fig.~\ref{fig:fig3}.

Fig.~\ref{fig:fig4}(c) features two peaks in the double-excitation probability as a function of inter-atomic distance. This highlights two separate excitation pathways. These peaks are associated to inter-atomic distances where $\delta = V_{ij}$ and $2\delta = V_{ij}$ which shows excitation through exciting $\ket{gr}$ or $\ket{rg}$ states and direct excitation to the $\ket{rr}$ state, respectively. Increasing the pulse duration gives us enough spectral resolution to resolve these two separate excitation pathways.

The effect of $T_{FWHM}$ becomes apparent when one looks at Fig.~4(d). For
small $T_{FWHM}$, there is a clear $P_1$ dominance. At these short pulse duration values the pump fields are broad enough to directly excite single Rydberg excitations without allowing significant double excitations. For a
longer duration of the pump pulses $P_2$ will dominate and $P_1$ will rapidly
fall with an onset of non-classical statistics around $0.3~\mu s$. As we mentioned earlier, with fixed peak field amplitudes the pump pulse area and spectral overlap of the pump with atoms at various distances become competing factors to determine the optimal pulse duration. These factors will
yield a continually higher $P_2/P_1^2$ as $T_{FWHM}$ increases, after
which the shrinking width of the anti-blockade will overcome the increasing
amplitude and cause $P_2/P_1^2$ to shrink. It is also optimal to set
the $T_{FWHM}$ to correspond to this optimal value of the ratio
$P_2/P_1^2$, even if it is at the expense of a lower $P_2$ value.

We also considered the probability of exciting the intermediate state, $\ket{e}$, in the two-photon excitation process. Given the single-photon detuning of $\Delta=100$~MHz, $P_e$ appears to remain constant. In general, we would like to avoid populating the intermediate state. This is to avoid noise at a potential retrieval where the Rydberg excitations are converted back to photons through $\ket{e}\rightarrow\ket{g}$ transition by re-applying the control field ($\Omega_2$). This will help to confirm generation of non-classical Rydberg excitation statistics by measuring auto-correlation values of recovered photons~\cite{Cantu2019}.

\begin{figure}[ht]
\centering
\includegraphics*[viewport =50 490 490 700,width=1\columnwidth]{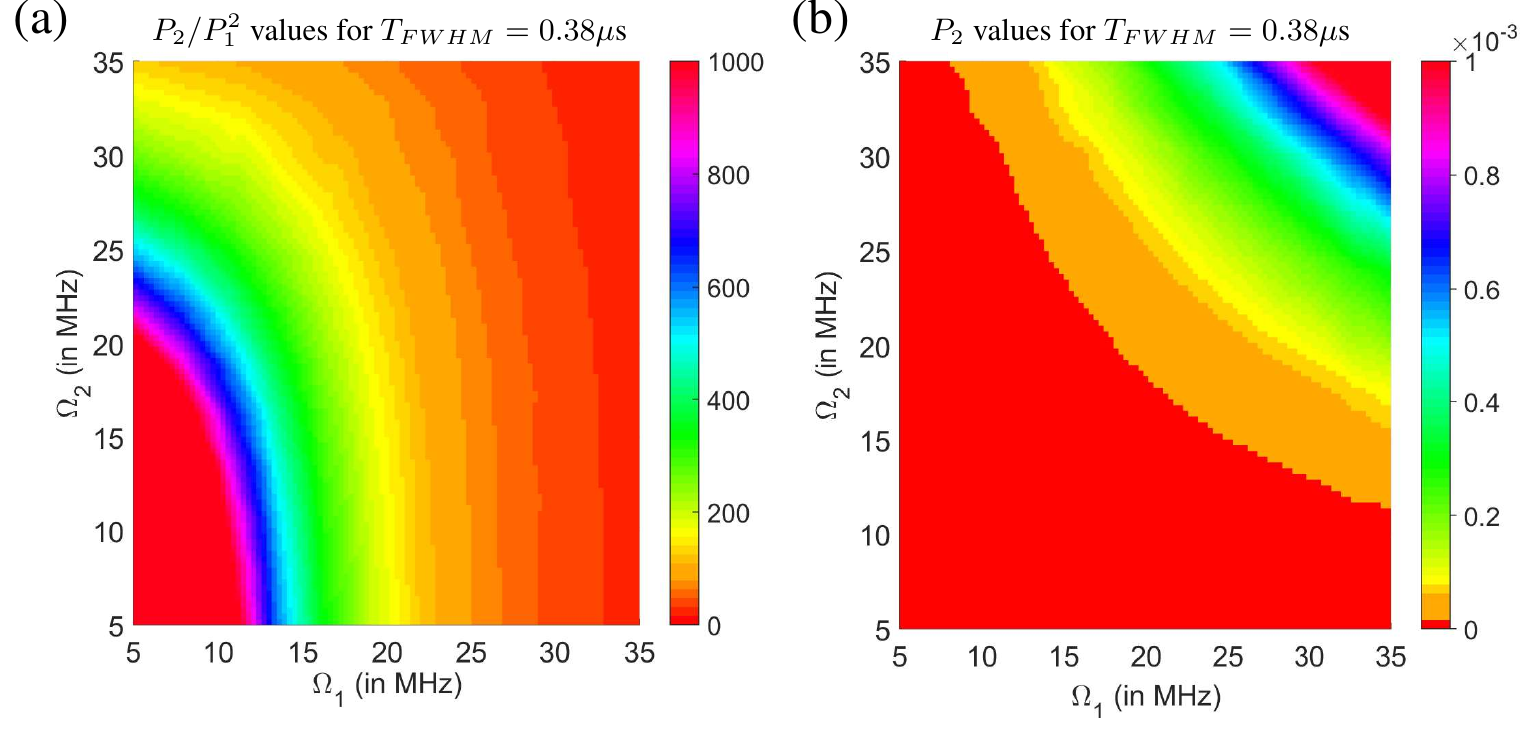}
\caption{(Color online) Effect of pump amplitudes in terms of $\Omega_1$, $\Omega_2$. (a)
Demonstrates the unequal effect the different amplitudes have on the ratio,
where both yield superior ratios for smaller values, yet as can be seen by the
yellow region for high $\Omega_2$ that the effect of $\Omega_1$ dominates. (b) Shows that increasing pump intensity for both fields has a similar impact on the probability of generating a doubly excited state. An effective coupling of form $\Omega = \Omega_1\Omega_2/\Delta$ can be used to describe coupling strength to the Rydberg state of an individual atom.}
\label{fig:fig5}
\end{figure}

Finding optimal experimental conditions to reach non-classical Rydberg excitation statistics is not limited to pulse duration. In particular, optimizing the excitation probability involves finding optimal values for pump field intensities. For both $\Omega_1$ and
$\Omega_2$, increasing the amplitude decreases $P_2/P_1^2$ ratios and increases the excitation probability, $P_2$. These are competing requirements, however, as it can be seen in Fig.~\ref{fig:fig5}, for any given $P_2/P_1^2$ ratio, one can find optimal Rabi frequency values that maximizes the excitation probability, $P_2$. The effect of pump field amplitudes on the ratio are not symmetric. The optimal range is such that $\Omega_2 > \Omega_1$,
with a maximized $\Omega_2$. To achieve a high $P_2$ along with a high
$P_2/P_1^2$, $\Omega_1$ must be decreased, while increasing $\Omega_2$. 

\subsection{Electromagnetically induced transparency}
In this subsection we focus on two-photon excitation dynamics through resonant coupling to the intermediate state, $\ket{e}$. This is often referred to as
Electromagnetically induced transparency (EIT) and has been experimentally demonstrated with
two-photon coupling to Rydberg states of atomic ensembles~\cite{Baur2014}. By
turning the control field ($\Omega_2$) off adiabatically, population can be
transferred from ground to the Rydberg state of the atoms. We achieve this using the error function for the temporal profile of the control field ($\Omega_2$) combined with a Gaussian pulse for the first pump field denoted by $\Omega_1$. This will allow us to adiabatically turn off the control field that will result in generating excitation in the Rydberg states of the atomic ensemble. For our numerical results, we use $\Omega_2(t)=A_2*\left[\text{erf}\left(-(t-T_{FWHM})*\sqrt{\theta_t})*100\right)+1\right]/2$ along with a Gaussian pulse for the first pump field ($\Omega_1$).
Below, we search for optimal conditions for the Rydberg anti-blockade under
resonant coupling; see Fig.~\ref{fig:fig1}(b).

\begin{figure}[ht]
\centering
\includegraphics*[viewport=50 270 550 705, width=1\linewidth]{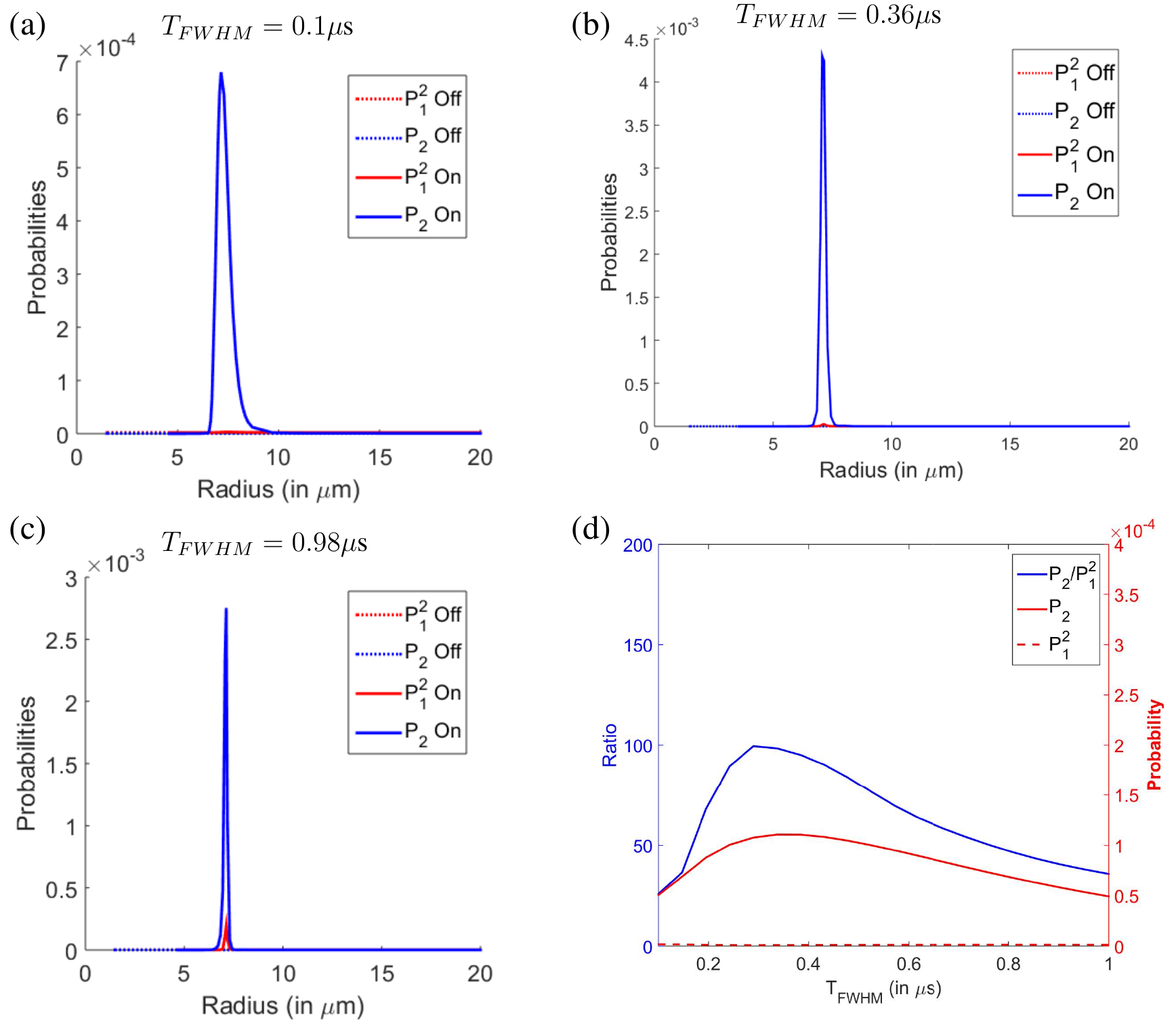}
\caption{(Color online) Effect of pulse duration in Rydberg EIT under Rydberg anti-blockade condition. The above results are based on $\Omega_1=14.4$~MHz, $\delta=50$~MHz, $\Omega_2=21.4$~MHz, and $\Delta=0$ in
which the Rydberg levels that are involved are n=75S. In (a) to (c) we
see increasing of peak value decreasing of width of the anti-blockade similar to the results under Raman coupling. (d) At $T_{FWHM}=0.36\mu s$ the
$\frac{P_2}{P_1^2}$ ratio is optimized.} 
\label{fig:fig6}
\end{figure}
The effect of increasing the pulse duration, $T_{FWHM}$ is similar to that of Raman coupling. As shown in Fig.~\ref{fig:fig6}(a--c) the longer pulse duration results in higher peak probability of excitation at the optimal distance. However this is accompanied by a narrower range (shell) of atoms contributing to the doubly excited state. The optimal pulse duration is about 0.36~$\mu$s.

Fig.~\ref{fig:fig4}(c) shows a double peak feature that
is associated to different excitation pathways. This effect also appears under EIT conditions with significantly lower visibility than the Raman coupling case. This is due to two factors. First the detuning from the Rydberg state is $\delta=50$~MHz in results shown in Fig.~\ref{fig:fig6}. This means that the excitation pathway through exciting $\ket{gr}$ and $\ket{rg}$ states is not spectrally covered (accessible) with these pulse durations. Second, Peak Rabi frequencies of the pump pulses are slightly lower than the Raman results. This will limit the probability of off-resonant coupling to $\ket{gr}$ and $\ket{rg}$ states.

%\begin{figure}[H]
%\centering
%\includegraphics*[width=\linewidth]{EITRatiovsTFWHM.png}
%	\label{fig:fig7a}
%\caption{\textbf{EIT $\mathbf{T_{FWHM}}$'s Effect on the Ratio}
%$\Omega_2=2.92MHz$, $\delta=10.36MHz$, $T_{FWHM}=2.177\mu s$, 70S interaction
%level, and linewidth $\gamma_e=2\pi\times6MHz$, the same setup as the previous
%previous figure. The optimal value of $T_{FWHM}=2.4\mu s$ shown in the
%previous figure is the value of $T_{FWHM}$ for which the ratio is maximized.}
%\label{fig:fig7}
%\end{figure}

\begin{figure}[ht]
\centering
\includegraphics*[viewport =50 490 490 700,width=1.0\columnwidth]{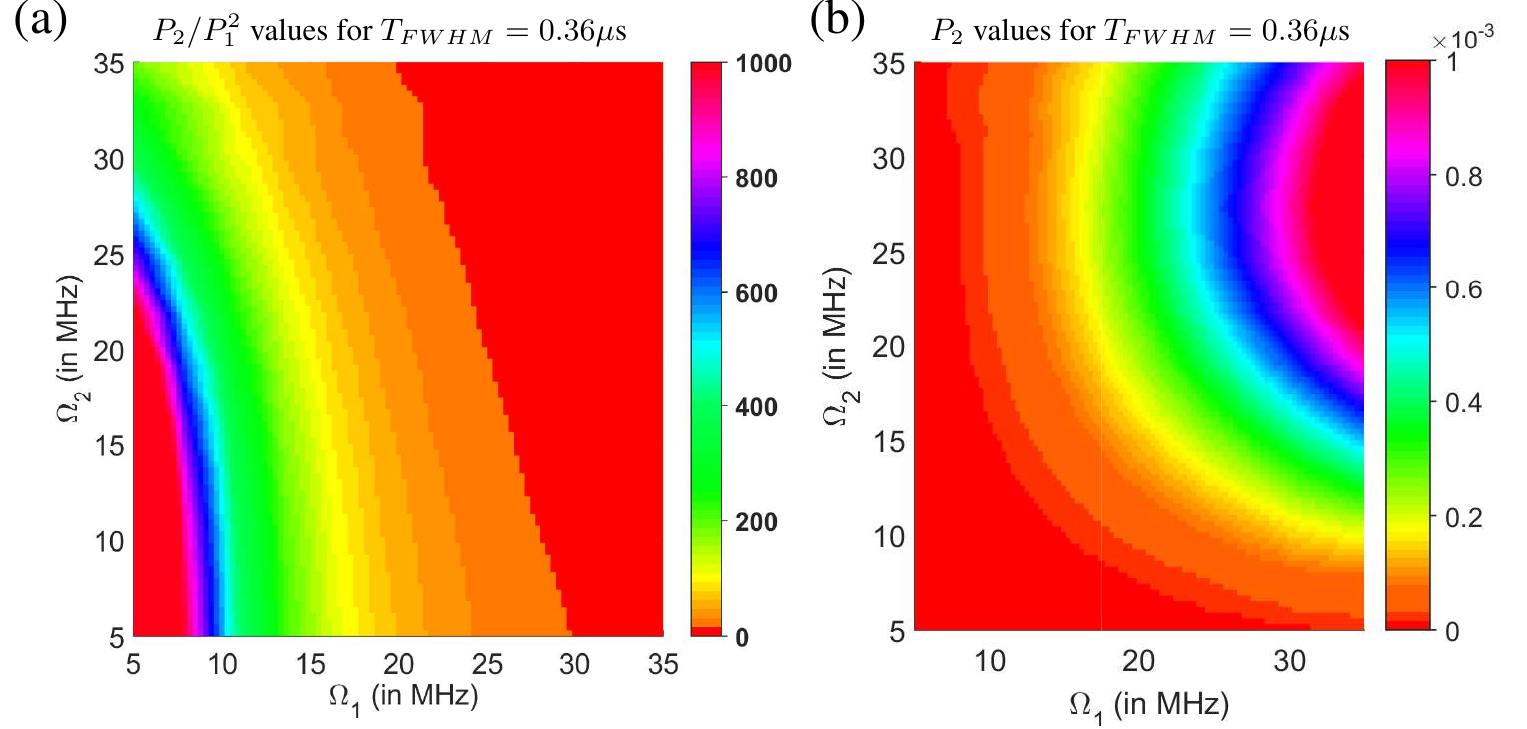}
\caption{(Color online) Effects of the Amplitude. In regards to $P_2/P_1^2$ (a) shows
a similar relation as in Raman, however with a more equal effect caused by
$\Omega_1$ and $\Omega_2$. In (b) the largest probability occurs at
$\Omega_2\approx 27.1MHz$ with $P_2\approx 0.0012$.}
\label{fig:fig7}
\end{figure}

Mapping $P_2$ and $\frac{P_2}{P_1^2}$ ratio as a function of $\Omega_1$ and
$\Omega_2$ allows us to find the optical condition for Rydberg anti-blockade
effect to result in a high-purity two-atom excitation under resonant coupling.
Similar to the previous case, shown in Fig.~\ref{fig:fig5}, there is a trade-off between
probability of generation ($P_2$) and purity ($P_2/P_1^2$). However, for EIT the effect of the amplitudes of the Rabi frequencies is significantly
different to that of Raman. As it can be seen in Fig.~\ref{fig:fig7}, the optimal conditions
can be reached at lower values for $\Omega_2$. This is in particular a more
favorable result as reaching large Rabi frequencies on the $\ket{e}\rightarrow\ket{r}$
transition can be challenging due to increasingly weak transition dipole
elements.

\begin{figure}[h]
\centering
\includegraphics*[viewport=50 490 500 700,width=\linewidth]{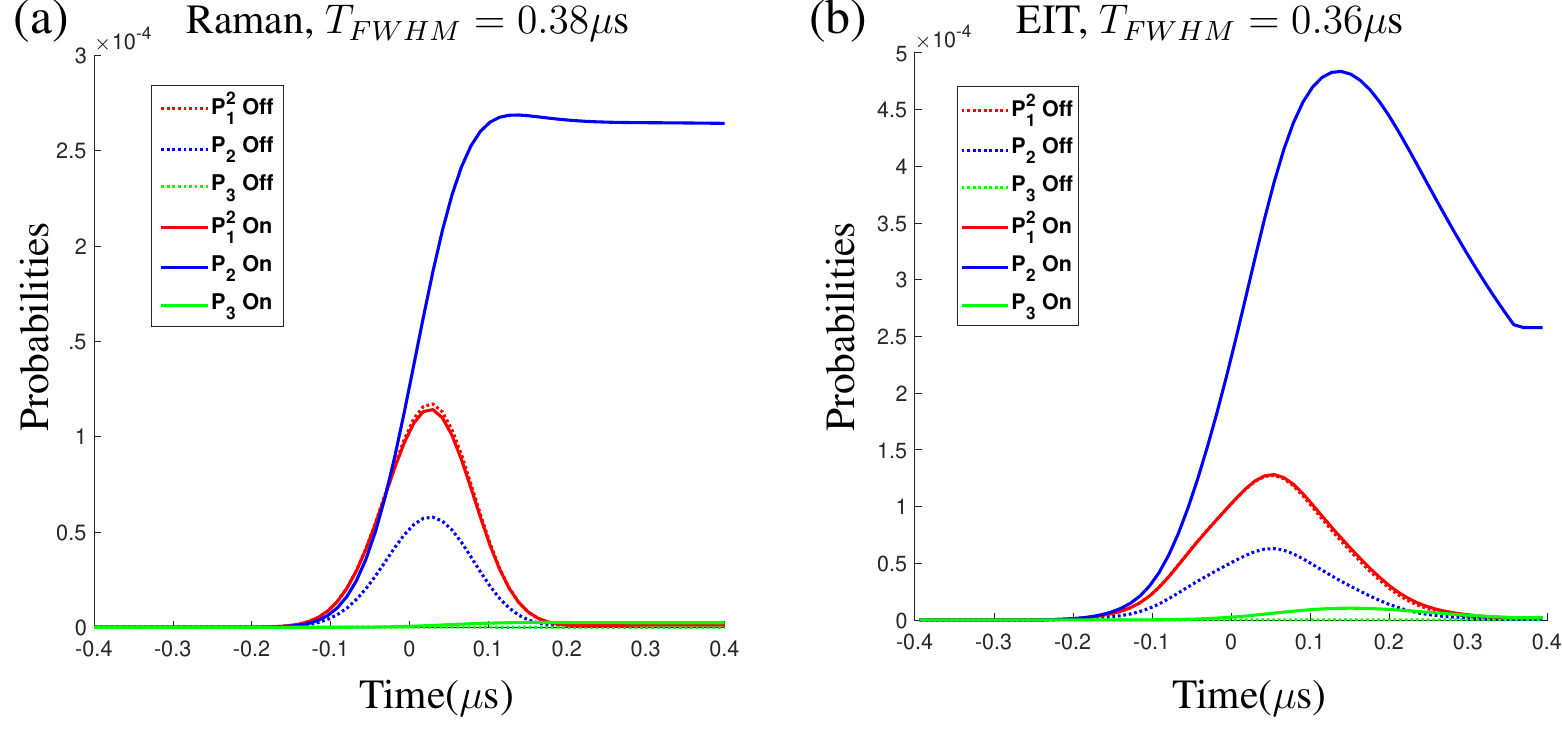}
\caption{(Color online) Three-atom dynamics. (a) Shows Rydberg excitation dynamics under
off-resonant Raman coupling with optimal parameters obtained in Fig.~\ref{fig:fig4}. (b)
Demonstrates results for Rydberg anti-blockade under the optimal conditions for
resonant coupling (EIT); see Fig.~\ref{fig:fig6}. These simulations use the grid integration
method described above with $P_{DF}$ shown in Fig.~\ref{fig:fig2}(b). }
\label{fig:fig8}
\end{figure}
\subsection{Higher number of Atoms}
One of the applications we envision for generating this type of atomic states
is to create a source of non-classical photons. Our results so far demonstrate
that Rydberg excitation dynamics can be modified to show highly non-classical
statistics. Another application of the control field, $\Omega_2$, would allow
us to convert these atomic excitations to photonic Fock states. However, given
that such conversion is inherently lossy, we need to quantify the probability
of exciting higher number of atoms and show that it remains limited. The
simulation of multi-atom Rydberg excitation dynamics can be performed for up to
about 10 atoms. However, sampling from all possible spatial configuration gets
increasingly difficult. In this subsection, we use the probability density
function shown in Fig.~\ref{fig:fig2}(b) and study three-atom Rydberg excitation dynamics;
see Fig.~\ref{fig:fig8}.

As demonstrated in Fig.~\ref{fig:fig8}, under both off-resonant and resonant pumping, the
probability of excitation three Rydberg atoms, $P_3$, remains significantly
smaller than $P_2$. This ensures that high-purity Fock states of $n=2$ can be
reached in Rydberg atomic ensembles.

Using the same physical parameters found in two-atom simulations, Fig.~\ref{fig:fig8}(a) and
(b) show that probability of three-atom excitation of $P_3=2.34\times10^{-6}$ and
$P_3=2.02\times10^{-6}$, respectively. These values are slightly higher than the $P_3$
values of a non-interacting system under similar conditions. However, in both
cases, $P_3$ is about two orders of magnitude smaller than the corresponding
two-atom excitation probabilities.

%\subsection*{Photon Output}
%\noindent\textbf{Wigner Function, Non-classicality and g2}
%The importance of the ratio $\frac{P_2}{P_1^2}$ is evident if one uses it to
%represent an emission spectrum.
%$$g_2(0)=\frac{\sum_{n} n(n-1)P_n}{\left[\sum_{n}nP_n\right]^2}$$
%$$g_2(0)\approx\frac{2P_2}{P_1^2}$$ As the number of atoms grows
%$\frac{P_2}{P_1^2}\rightarrow 1/2$, one will then have a coherent state.

%The existence of a negative value in the Wigner Function implies
%non-classicality. Such an event can not be observed within the two atom case,
%due to the dominance of vacuum ($P_0$). However it should be able to be
%observed for larger number of atoms due to scaling. In the N atom case, where
%...  

%\noindent\textbf{Template Title} 
 %For most of the results $R_{FWHM}$ will be taken to infinity, or a flat pump in the space domain will be used.
\
\section{Discussion}
Predicting multi-atom excitation dynamics in an interacting system of Rydberg
atoms is intractable. Considering random positions of atoms in an atomic cloud
adds another layer of complication. Different approaches have been taken to
predict regimes of non-classical Rydberg excitation
statistics~\cite{Stanojevic2010,Wuster2010,Ates2007} which led to experimental
observation of the Rydberg anti-blockade effect~\cite{Amthor2010,Viteau2012} and quantum nonlinear optics~\cite{Cantu2019,Sinclair2019}.
In this work, we explored the effect of experimentally controlled parameters
such as pulse duration, detuning, and amplitudes, to find the optimal
conditions leading to highly non-classical Rydberg excitation statistics. We
numerically studied two-atom and three-atom Rydberg excitation dynamics
averaged over many possible spatial configurations for a could of 1000 atoms in
a dipole trap. Our approach allowed us to consider spatial effects associated
to the density of atoms and spatial variations in the driving fields. The
effect of atomic density manifests itself indirectly by finding the
\emph{right} Rydberg level and two-photon detuning that will optimize the
Rydberg anti-blockade effect. We also explored spatial variations of the pump, for example, due
to using orbital angular momentum modes of light. The results are not explicitly discussed here in the manuscript as we observe no
significant impact on the outcomes such as Rydberg excitation probabilities.
This is primarily due to random positioning of the atoms in the cloud.

We show that one can achieve non-classical Rydberg excitation statistics with
auto-correlation values up to 1000 using both off-resonant and resonant driving
fields. Our approach can be used to guide experimental efforts in demonstrating
such Rydberg excitation dynamics and to convert these excitations to generate
photonic Fock states.

{\it Acknowledgement} -- JS and KB-F thank Hudson Pimenta for valuable discussions. This work was supported by the Natural Sciences and Engineering Research Council of Canada (NSERC) through its  Discovery  Grant and National Research Council's student program.

%Bibliography

%
% Acknowledgements
%
\vspace{0.2cm}
%\noindent\textbf{Author Contributions}
\noindent 
\vspace{0.5 EM}

%\noindent\textbf{Acknowledgments} This work was supported by Natural Sciences and Engineering Research Council of Canada (NSERC). 

%\vspace{0.5 EM}

%\noindent\textbf{Author Information}
%\noindent Correspondence and requests for materials should be addressed to jr4taylo@uwaterloo.ca and khabat.heshami@nrc-cnrc.gc.ca.

%\vspace{0.5 EM}

%\noindent\textbf{Competing interests} The authors declare no competing financial interests.

%\vspace{0.5 EM}

%\noindent\textbf{Data availability statement}
%The data that support the plots and analysis within this paper and other findings of this study are available from the corresponding author upon reasonable request.
%\section*{Appendix A}
%\subsection*{Higher $\frac{P_2}{P_1^2}$}

%\begin{figure}[h]
%\centering

%\includegraphics*[viewport=0 500 520 720, %width=\linewidth]{figA1.pdf}
%\caption{EIT Achieving $\frac{P_2}{P_1^2}=1000$. (a) Demonstrates that when aiming for a higher ratio $T_{FWHM}$'s effect is unchanged, with approximately the same peak. (b) Time dynamics for $\Omega_1=7.3MHz$, $\Omega_2=17.3MHz$, $T_{FWHM}=0.48\mu s$, achieves $\frac{P_2}{P_1^2}= 1.0315E+03$ and $P_2=7.165E-6$. }
%\label{fig:figA1}
%\end{figure}

%\begin{figure}[h]
%\includegraphics*[viewport=0 500 520 720, width, width=\linewidth]{figA2.pdf}
%\caption{Raman Achieving $\frac{P_2}{P_1^2}=1000$ (a) Demonstrates that when aiming for a higher ratio $T_{FWHM}$'s effect is unchanged. (b) Time dynamics for $\Omega_1=11.32MHz$, $\Omega_2=30MHz$, $T_{FWHM}=0.52\mu s$, achieves $\frac{P_2}{P_1^2}=926$ and $P_2=1.377E-5$. }
\label{fig:figA2}
%\end{figure}

%\noindent\textbf{Space Domain, and Other Modes} 
%As described before there are multiple different forms $\Omega(r,t)$ can take. The simulation was ran for different possible Laguerre Guass modes of the pump laser's intensity, here $T_{FWHM}$ and $R_{FWHM}$ are still used to represent the characteristic values. The same Omega equation applies as stated before but with a different $I(r,t)$, below is the form $I(r,t)$ takes where $\theta_r=\frac{4\ln(2)}{R_{FWHM}^2}$ and $\theta_t=\frac{4\ln(2)}{T_{FWHM}^2}$:
%$$I(r,t)=|A\left(\sqrt{\theta_r r}\right)^l*P_{p}^l(\theta_r r^2)*\exp(-\theta_r r^2)*cos(l*\theta)\exp(-\theta_t t^2)|$$

%The simulation for the two level system was ran for $p,l =(0,0),(0,1),(1,0),(1,1)$ with multiple values of $R_{FWMH}$ and little to no improvement was found for $\frac{P_2}{P_1^2}$. However when $R_{FWHM}\neq\infty$, require far more computation power, thus the results focus on $R_{FWHM}=\infty$ with the only non-zero mode which is $p=0,l=0$.


\begin{thebibliography}{10}
\bibitem{ChangRev2014} D. E. Chang, V. Vuleti\'c, and M. D. Lukin, Nature Photonics {\bf 8}, 685 (2014).
\bibitem{SaffmanRev} M. Saffman, T. G. Walker, and K. M{\o}lmer, Rev. Mod. Phys. {\bf 82}, 2313 (2010).
\bibitem{SibalicAdamsBook} Nikola \v{S}ibali\'c, Charles S. Adams. "Rydberg physics." ISBN:. IOP ebooks. Bristol, UK: IOP Publishing (2018).
\bibitem{ParedesBarato2014} D. Paredes-Barato and C. S. Adams, Phys. Rev. Lett. {\bf 112}, 040501 (2014).
\bibitem{Khazali2015} M. Khazali, K. Heshami, and C. Simon, Physical Review A {\bf 91}, 030301 (2015).
\bibitem{Saffman2015} K. M. Maller, M. T. Lichtman, T. Xia, Y. Sun, M. J. Piotrowicz, A. W. Carr, L. Isenhower, and M. Saffman
Phys. Rev. A {\bf 92}, 022336 (2015).
\bibitem{Tiarks2019} D. Tiarks, S. Schmidt-Eberle, T. Stolz, G. Rempe, and S. D{\"u}rr, Nature Physics {\bf 15}, 124--126 (2019).
\bibitem{Lukin2018} H. Levine, A. Keesling, A. Omran, H. Bernien, S. Schwartz, A. S. Zibrov, M. Endres, M. Greiner, V. Vuleti\'{c}, and M. D. Lukin, Phys. Rev. Lett. {\bf 121}, 123603 (2018).
\bibitem{Ding2019} D.-S. Ding, Y.-C. Yu, M.-X. Dong, Y.-H. Ye, G.-C. Guo, B.-S. Shi, arXiv:1903.08303 (2019).

\bibitem{Weimer2010} H. Weimer, M. M\"{u}ller, I. Lesanovsky, P. Zoller, and H. P. B\"{u}chler, Nature Physics {\bf 6}, 382–388 (2010).
\bibitem{Bernien2017} H. Bernien, S. Schwartz, A. Keesling, H. Levine, A. Omran, H. Pichler, S. Choi, A. S. Zibrov, M. Endres, M. Greiner, V. Vuletić, and M. D. Lukin, Nature {\bf 551}, 579–584 (2017).
\bibitem{Keesling2019} A. Keesling, A. Omran, H. Levine, H. Bernien, H. Pichler, S. Choi, R. Samajdar, S. Schwartz, P. Silvi, S. Sachdev, P. Zoller, M. Endres, M. Greiner, V Vuleti\'{c}, and M. D. Lukin, Nature {\bf 568}, 207–211 (2019).
\bibitem{Montenegro2013} M. M. M\"{u}ller, A. K\"{o}lle, R. L\"{o}w, T. Pfau, T. Calarco, and S. Montangero, Phys. Rev. A {\bf 87}, 053412 (2013).
\bibitem{Craddock2019} A.N. Craddock, J. Hannegan, D.P. Ornelas-Huerta, J.D. Siverns, A.J. Hachtel, E.A. Goldschmidt, J.V. Porto, Q. Quraishi, and S.L. Rolston, Phys. Rev. Lett. {\bf 123}, 213601 (2019).
\bibitem{Khazali2017} M. Khazali, K. Heshami, C. Simon, Journal of Physics B: Atomic, Molecular and Optical Physics {\bf 50}, p.215301 (2017).
\bibitem{Peyronel2012} T. Peyronel, O. Firstenberg, Q.-Y. Liang, S. Hofferberth, A. V. Gorshkov, T. Pohl, M. D. Lukin, and V. Vuleti\'{c}, Nature {\bf 488}, 57–60 (2012).
\bibitem{Gorshkov2011} A. V. Gorshkov, J. Otterbach, M. Fleischhauer, T. Pohl, and M. D. Lukin, Phys. Rev. Lett. {\bf 107}, 133602 (2011).
\bibitem{Hoferberth2014} H. Gorniaczyk, C. Tresp, J. Schmidt, H. Fedder, and S. Hofferberth, Phys. Rev. Lett. {\bf 113}, 053601 (2014).
\bibitem{Kuzmich2012} Y. O. Dudin, A. Kuzmich, Science {\bf 336}, pp. 887--889 (2012).
\bibitem{He2014} B. He, A. V. Sharypov, J. Sheng, C. Simon, and M. Xiao, Phys. Rev. Lett. {\bf 112}, 133606 (2014).
\bibitem{Cantu2019} S. H. Cantu, A. V. Venkatramani, W. Xu, L. Zhou, B. Jelenkovi\'c, M. D. Lukin, V. Vuleti\'c, arXiv:1911.02586 (2019).
\bibitem{Sinclair2019} J. Sinclair, D. Angulo, N. Lupu-Gladstein, K. Bonsma-Fisher, A. M. Steinberg, arXiv:1906.05151 (2019).
\bibitem{Han2010} Y. Han, B. He, K. Heshami, C.-Z. Li, C. Simon, Physical Review A {\bf 81}, 052311 (2010).
\bibitem{Stanojevic2010} J. Stanojevic and R. C\^ot\'e, Phys. Rev. A {\bf 81}, 053406 (2010).
\bibitem{Wuster2010} S. W\"uster, J. Stanojevic, C. Ates, T. Pohl, P. Deuar, J. F. Corney, and J. M. Rost, Phys. Rev. A {\bf 81}, 023406 (2010).
\bibitem{Ates2007} C. Ates, T. Pohl, T. Pattard, and J. M. Rost, Phys. Rev. Lett. {\bf 98}, 023002 (2007).
\bibitem{Amthor2010}T. Amthor, C. Giese, C. S. Hofmann, and M. Weidemüller, Phys. Rev. Lett. {\bf 104}, 013001 (2010).
\bibitem{Viteau2012} M. Viteau, P. Huillery, M. G. Bason, N. Malossi, D. Ciampini, O. Morsch, E. Arimondo, D. Comparat, and P. Pillet, Phys. Rev. Lett. {\bf 109}, 053002 (2012).
\bibitem{Baur2014} S. Baur, D. Tiarks, G. Rempe, and S. D\"urr, Phys. Rev. Lett. {\bf 112}, 073901 (2014).
%\bibitem{AT1955} S. H., Autler, C. H. Townes, Stark effect in rapidly varying fields. Phys. Rev. {\bf 100}, 703 (1955).
%\bibitem{Saglamyurek2018} E. Saglamyurek, T. Hrushevskyi, A. Rastogi, K. Heshami, and L. J. LeBlanc, Nature Photonics {\bf 12}, 774 (2018).

\end{thebibliography}
\end{document}